\documentclass[aps,preprint,amsmath,amssymb]{revtex4}
\usepackage{graphicx}
\begin{document}

\title{Probe of the electromagnetic moments of the tau lepton in gamma-gamma collisions
at the CLIC}

\author{A. A. Billur}
\email[]{abillur@cumhuriyet.edu.tr} \affiliation{Department of
Physics, Cumhuriyet University, 58140, Sivas, Turkey}

\author{M. K\"{o}ksal}
\email[]{mkoksal@cumhuriyet.edu.tr} \affiliation{Department of
Physics, Cumhuriyet University, 58140, Sivas, Turkey}

\begin{abstract}
We have investigated the electromagnetic moments of the tau lepton in $e^{+}e^{-} \rightarrow e^{+}\gamma^{*} \gamma^{*} e^{-}\rightarrow e^{+}\tau \bar{\tau}e^{-}$ process at the CLIC.
We have obtained $95\%$ confidence level bounds on the anomalous  magnetic and electric dipole moments
for various values of the integrated luminosity and the center of mass energy.
We have shown that the $e^{+}e^{-} \rightarrow e^{+}\gamma^{*} \gamma^{*} e^{-}\rightarrow e^{+}\tau \bar{\tau}e^{-}$ process at the CLIC leads to a remarkable improvement in the existing
experimental bounds on the anomalous  magnetic and electric dipole moments.
\end{abstract}


\maketitle

\section{Introduction}

The Land\'{e} g-factor or gyromagnetic factor $g$ is described by the formula between particle's magnetic moment $\vec{\mu}$ and it's spin $\vec{s}$: $\vec{\mu}=g (\mu_{B} / \hbar)\vec{s}$
where $\mu_{B}$ is the Bohr magneton. In the Dirac equation, the value of $g$ is $2$ for a point-like particle. Deviation from this value $a=\frac{(g-2)}{2}$ is called as the anomalous magnetic moment, and without anomalous and radiative corrections $a=0$. However, the anomalous magnetic moment $a_{e}$ of the electron was firstly obtained using radiative corrections by Schwinger as $a_{e}=\frac{\alpha}{2\pi}=0.001161$ \cite{1}. So far, the accuracy of the $a_{e}$ was examined by many theoretical and experimental studies. These studies have provided the most precise determination of fine-structure constant $\alpha_{QED}$, since $a_{e}$ is quite senseless to the strong and weak interactions. On the other hand, the anomalous magnetic moment $a_{\mu}$ of the muon enables testing the Standart Model (SM) and investigating alternative theories to the SM. The $a_{e}$ and $a_{\mu}$ can be obtained with high sensitivity through spin precession experiment. Otherwise, the spin precession experiment cannot be used to measure the anomalous magnetic moment $a_{\tau}$ of the tau, because of the relatively short lifetime $2.906 \times 10^{-13}$ s of tau \cite{tau}. So the current bounds of the $a_{\tau}$ are obtained by collision experiments. The theoretical value of the $a_{\tau}$ from QED is given as $a^{SM}_{\tau}=0.001177 $ \cite{2,3}.

The experimental bounds on the the $a_{\tau}$ are provided by the L3: $-0.052<a_{\tau}<0.058$ , OPAL: $-0.068<a_{\tau}<0.065$ and DELPHI: $-0.052<a_{\tau}<0.013$ collaborations at the LEP at $95\%$ C.L. \cite{4,5,6}.

CP violation was firstly observed in a small fractions of $K_{L}^{0}$ mesons decaying to two pions in the SM \cite{7}. This phenomenology in the SM can be easily introduced by  the Cabibbo-Kobayashi-Maskawa mechanism in the quark sector \cite{8}. On the other hand, there is no CP violation in the lepton sector. However, CP violation in the quark sector causes a very small electric dipole moment of the leptons. At least to three-loop are required in order to produce a nonzero contributing in the SM and it's crude estimate is obtained as $|d_{\tau}|\leq 10^{-34}\, e\,cm$ \cite{9}.

If at least two of the three neutrinos have different mass values, CP violation in the lepton sector can occur as similar to the CP violation in the quark sector \cite{10}. There are many different models beyond the SM inducing to CP violation in the lepton sector. These models are leptoquark \cite{11,12}, SUSY \cite{13}, left-right symmetric \cite{14,bk} and more Higgs multiplets \cite{15,16}.

The bounds at $95\%$ C. L. on the anomalous electric dipole moment of the tau
yield by LEP experiments L3: $|d_{\tau}|<3.1 \times 10^{-16}\, e\,cm$, OPAL: $|d_{\tau}|<3.7 \times 10^{-16}\, e\,cm$,  and DELPHI: $|d_{\tau}|<3.7 \times 10^{-16}\, e\,cm$. The most restrictive experimental bounds are obtained by BELLE: $-2.2<Re(d_{\tau})<4.5 \times (10^{-17}\, e\,cm)$ and
$-2.5<Im(d_{\tau})<0.8 \times (10^{-17}\, e\,cm)$. There are model dependent and independent studies on the anomalous dipole moments of the tau lepton in the literature \cite{47,phe1,phe2,phe3,phe4,phe5,phe7,phe8,phe9}.

We consider that difference between $a_{\tau}^{SM}$ ($d_{\tau}^{SM}$) and $a_{\tau}^{exp}$ ($d_{\tau}^{exp}$) can be reduced to determine precisely a new term proportional to $F_{2}$ ($F_{3}$) to the SM $\tau\tau\gamma$ vertex.
For this reason, the electromagnetic vertex factor of the tau lepton can be parameterized

\begin{eqnarray}
\Gamma^{\nu}=F_{1}(q^{2})\gamma^{\nu}+\frac{i}{2 m_{\tau}}F_{2}(q^{2}) \sigma^{\nu\mu}q_{\mu}+\frac{1}{2 m_{\tau}}F_{3}(q^{2}) \sigma^{\nu\mu}q_{\mu}\gamma^{5}
\end{eqnarray}
where $\sigma_{\nu\mu}=\frac{i}{2}(\gamma_{\nu}\gamma_{\mu}-\gamma_{\mu}\gamma_{\nu})$, $q$ is the momentum transfer to the photon and $m_{\tau}=1.777$ GeV is the mass of tau lepton. In the SM, at tree level, $F_{1}=1$, $F_{2}=0$ and $F_{3}=0$. Besides, in the loop effects arising from the SM and the new physics, $F_{2}$ and $F_{3}$ may be not equal to zero. For example, the anomalous coupling $F_{2}$ is given by

\begin{eqnarray}
F_{2}(0)=a_{\tau}^{SM}+a_{\tau}^{NP}
\end{eqnarray}
where $a_{\tau}^{SM}$ is the contribution of the SM and $a_{\tau}^{NP}$ is the contribution of the new physics \cite{c1,c2,c3,c4}.Therefore, the $q^{2}$-dependent form factors $F_{1}(q^{2}),F_{2}(q^{2})$ and $F_{3}(q^{2})$ in limit $q^{2} \rightarrow 0$ are given by,

\begin{eqnarray}
F_{1}(0)=1,\: F_{2}(0)=a_{\tau},\: F_{3}(0)=\frac{2m_{\tau}d_{\tau}}{e}.
\end{eqnarray}

The Compact Linear Collider (CLIC) is a proposed future $e^{+}e^{-}$ collider, designed to fulfill $e^{+}e^{-}$ collisions at energies from $0.5$ to $3$ TeV \cite{17}, and it is planned to be constructed with a three main stages research region \cite{18}. The CLIC has been extensively studied for interactions beyond the SM \cite{20,21,22,23,24,25,26,27,28,29,30,31,32,33,34,37,38}. The CLIC enables to investigate the $\gamma\gamma$ and $\gamma e$ interactions by converting the original $e^{-}$ or $e^{+}$ beam into a photon beam through the laser backscattering procedure \cite{34x,35,36}. One of the other well-known applications of the CLIC is the $\gamma ^{*} \gamma^{*}$ process, where the emitted quasireal photon $\gamma^{*}$ is scattered with small angle from the beam pipe of $e^{-}$ or $e^{+}$ \cite{39,40,41,42,pes}. Since these photons have a low virtuality ($Q_{max}^{2}=2$\,GeV$^{2}$), they are almost on mass shell. $\gamma ^{*} \gamma^{*}$ processes can be described by equivalent photon approximation, i.e. using the Weizsacker-Williams approximation \cite{43,44,45,46,47,48,49,50,51,52,53,54,app}. Such processes have experimentally observed at the LEP, Tevatron and LHC \cite{d1,d2,d3,d4,d5,d6,d7}.
There are two reasons why we have chosen the CLIC in this work:
First, the observation of the most stringent experimental bound on the anomalous magnetic dipole moment of the tau lepton by using multiperipheral collision at the LEP through the process $e^{+}e^{-} \rightarrow e^{+}\tau \bar{\tau}e^{-}$  \cite{6}. Secondly, the importace of high center-of-mass energies to examine the electromagnetic properties of tau lepton since  anomalous $\tau\tau\gamma$ couplings depend on more energy than SM $\tau\tau\gamma$ couplings at the tree level. Therefore, we investigate the potential of CLIC via the process $e^{+}e^{-} \rightarrow e^{+}\tau \bar{\tau}e^{-}$  to examine the anomalous  magnetic and electric dipole moments of tau lepton.

\section{Cross sections and numerical analysis}

During calculations, the CompHEP-$4.5.1$ program was used by including the new interaction vertices \cite{55}. Also, the acceptance cuts were imposed as $|\eta_{\tau}|<2.5$ for  pseudorapidity, $p_{T}^{\tau}>20 \: $ GeV for transverse momentum cut of the final state particles, $\Delta R_{\tau \bar{\tau}}>0.5$ the separation of final tau leptons.

We show the integrated total cross-section of the process
$e^{+}e^{-} \rightarrow e^{+}\gamma^{*} \gamma^{*} e^{-}\rightarrow e^{+}\tau \bar{\tau}e^{-}$ as a function of the anomalous couplings $F_{2}$ and $F_{3}$ in Fig. $1$ for three different center-of-mass energies. As can be seen in Fig. $1$, while the total cross section is symmetric for anomalous coupling $F_{3}$, it is nonsymmetric for $F_{2}$.

We estimate $95\%$ C. L. bounds on anomalous coupling parameters $F_{2}$ and $F_{3}$ using $\chi^{2}$ test. The $\chi^{2}$ function is described by the following formula

\begin{eqnarray}
\chi^{2}=\left(\frac{\sigma_{SM}-\sigma(F_{2},F_{3})}{\sigma_{SM}\delta}\right)^{2},
\end{eqnarray}
where $\delta=\sqrt{(\delta_{st})^{2}+(\delta_{sys})^{2}}$; $\delta_{st}=\frac{1}{\sqrt{N_{SM}}}$ is the statistical error and $\delta_{sys}$ is the systematic error.
The number of expected events is calculated as the signal $N=L_{int}\times BR \times \sigma$ where
$L_{int}$ is the integrated luminosity. The tau lepton decays roughly $35\%$ of the time leptonically and $65\%$ of the time to one or more hadrons. So we consider one of the tau leptons decays leptonically and the other hadronically for the signal. Thereby, we assume that branching ratio of the tau pairs in the final state to be $BR=0.46$.

There are systematic uncertainties in exclusive production at the lepton and hadron colliders. For the process $e^{+}e^{-} \rightarrow e^{+}e^{-}\tau^{+}\tau^{-}$, systematic errors are experimentally studied between $4.3\%$ and $9\%$ at the LEP
\cite{6,55x}. Recently, exclusive lepton production at the LHC has been examined and its systematic uncertainty is $4.8\%$ \cite{d4}. Also, the process $p p \rightarrow p\tau^{+}\tau^{-}p$ with  $2\%$ of the total systematic error at the LHC has investigated phenomenologically in Ref. [19]. Therefore, the sensitivity limits on the anomalous magnetic and electric dipole moments of the tau lepton through the process $e^{+}e^{-} \rightarrow e^{+}e^{-}\tau^{+}\tau^{-}$ have calculated by considering three systematic errors: $2\%$, $5\%$ and $10\%$. On the other hand, there may occur an uncertainty arising from virtuality of $\gamma^{*}$ used in the Weizsacker-Williams approximation.
In Figs. $2$-$4$, we have calculated the integrated cross sections as a function of $F_{2}$ and $F_{3}$ for different $Q_{max}^{2}$ values. We can see from these figures the total cross section changes slightly with the variation of the $Q_{max}^{2}$ value. The sensitivity limits on the anomalous couplings $a_{\tau}$ and $d_{\tau}$ for different values of photon virtuality, center-of-mass energy and luminosity has been given in Table I.  It has shown that the bounds on the anomalous couplings do not virtually change when $Q_{max}^{2}$ increases. Therefore, we can understand that the large values of $Q_{max}^{2}$ do not bring an important contribution to obtain sensitivity limits on the anomalous couplings \cite{4,5,51}.

In Tables II-IV, we show 95\% C.L. sensitivity bounds of the coupling $a_{\tau}$ and ${d_{\tau}}$
for various systematic uncertainties, integrated CLIC luminosities and center of mass energies. While calculating the table values, we assumed that at a given time, only one of the anomalous couplings deviated from the SM.
In Fig. $5$, we demonstrate the sensitivity contour plot at $95\%$ C.L. for the anomalous couplings $F_{2}$ and $F_{3}$ at the $\sqrt{s}=0.5$, $1.5$ and $3$ TeV with corresponding maximum luminosities through process $e^{+}e^{-} \rightarrow e^{+}\gamma^{*} \gamma^{*} e^{-}\rightarrow e^{+}\tau \bar{\tau}e^{-}$.

\section{Conclusions}

The CLIC as a $\gamma^{*} \gamma^{*}$ collider using the Weizsacker-Williams  virtual photon fields of the $e^{-}$ and $e^{+}$ provides an ideal venue to investigate the electromagnetic moments of the tau lepton. For this reason, we have studied the potential of $e^{+}e^{-} \rightarrow e^{+}\gamma^{*} \gamma^{*} e^{-}\rightarrow e^{+}\tau \bar{\tau}e^{-}$ at the CLIC to examine the anomalous magnetic and electric dipole moments of the tau lepton. The findings of this study show that the CLIC can improve the sensitivity bounds on anomalous couplings electromagnetic dipole moments of tau lepton with respect to the LEP bounds.

\pagebreak

\pagebreak

\begin{figure}
\includegraphics{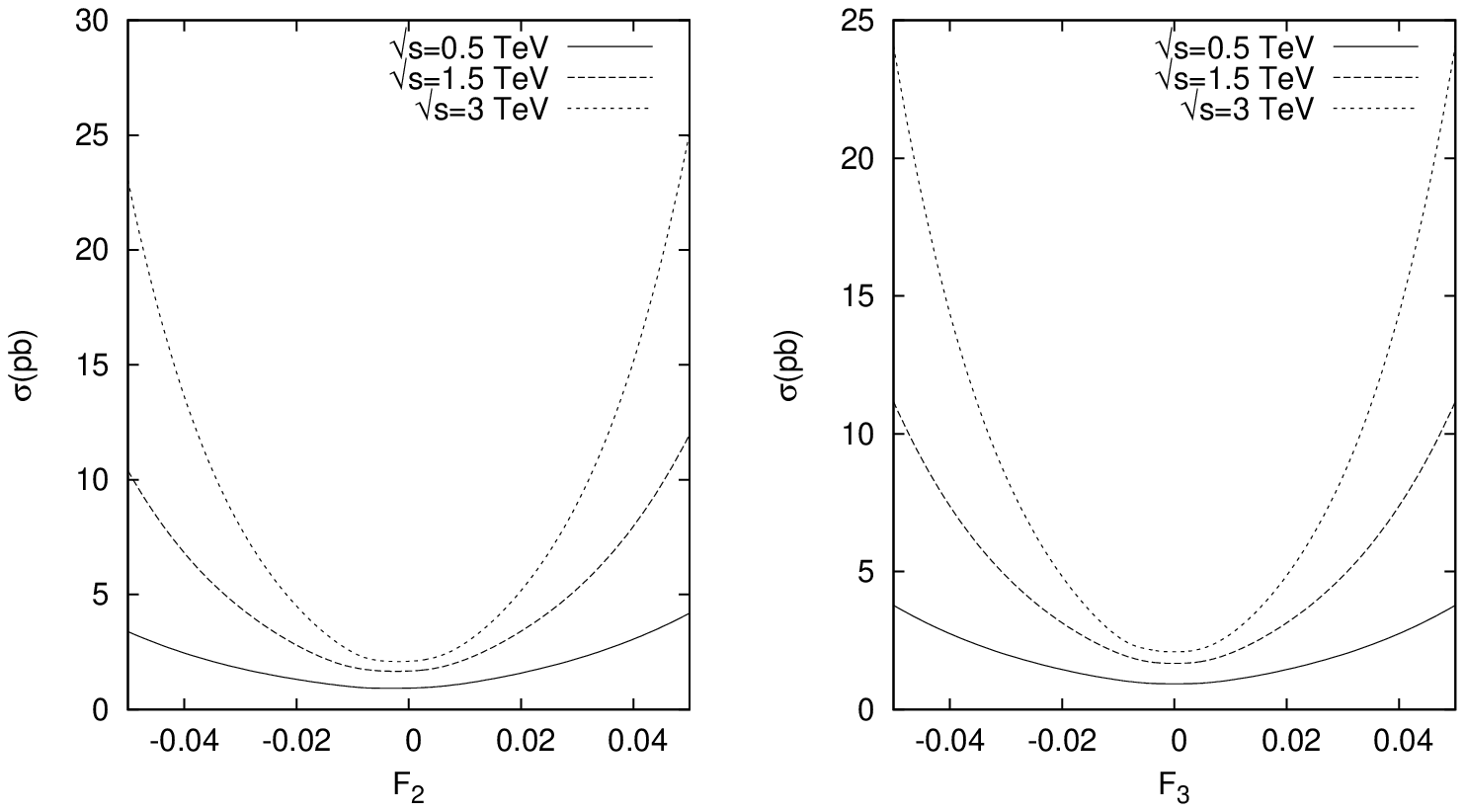}
\caption{The integrated total cross-section of the process $e^{+}e^{-} \rightarrow e^{+}\gamma^{*} \gamma^{*} e^{-}\rightarrow e^{+}\tau \bar{\tau}e^{-}$ as a function of anomalous couplings $F_{2}$ and $F_{3}$ for three different center-of-mass energies.
\label{fig1}}
\end{figure}

\begin{figure}
\includegraphics{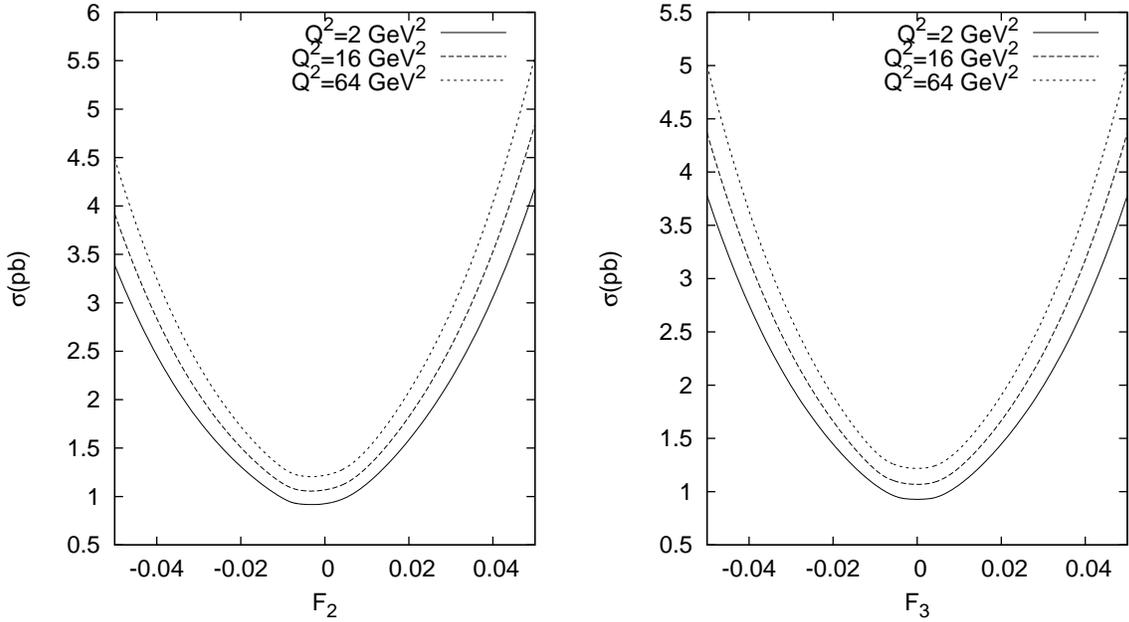}
\caption{The total cross section as a function of $F_{2}$ and $F_{3}$ for different values of $Q^{2}$ at the center of mass energy $\sqrt{s}=0.5$ TeV for the process $e^{+}e^{-} \rightarrow e^{+}\gamma^{*} \gamma^{*} e^{-}\rightarrow e^{+}\tau \bar{\tau}e^{-}$.
\label{fig2}}Fig.
\end{figure}

\begin{figure}
\includegraphics{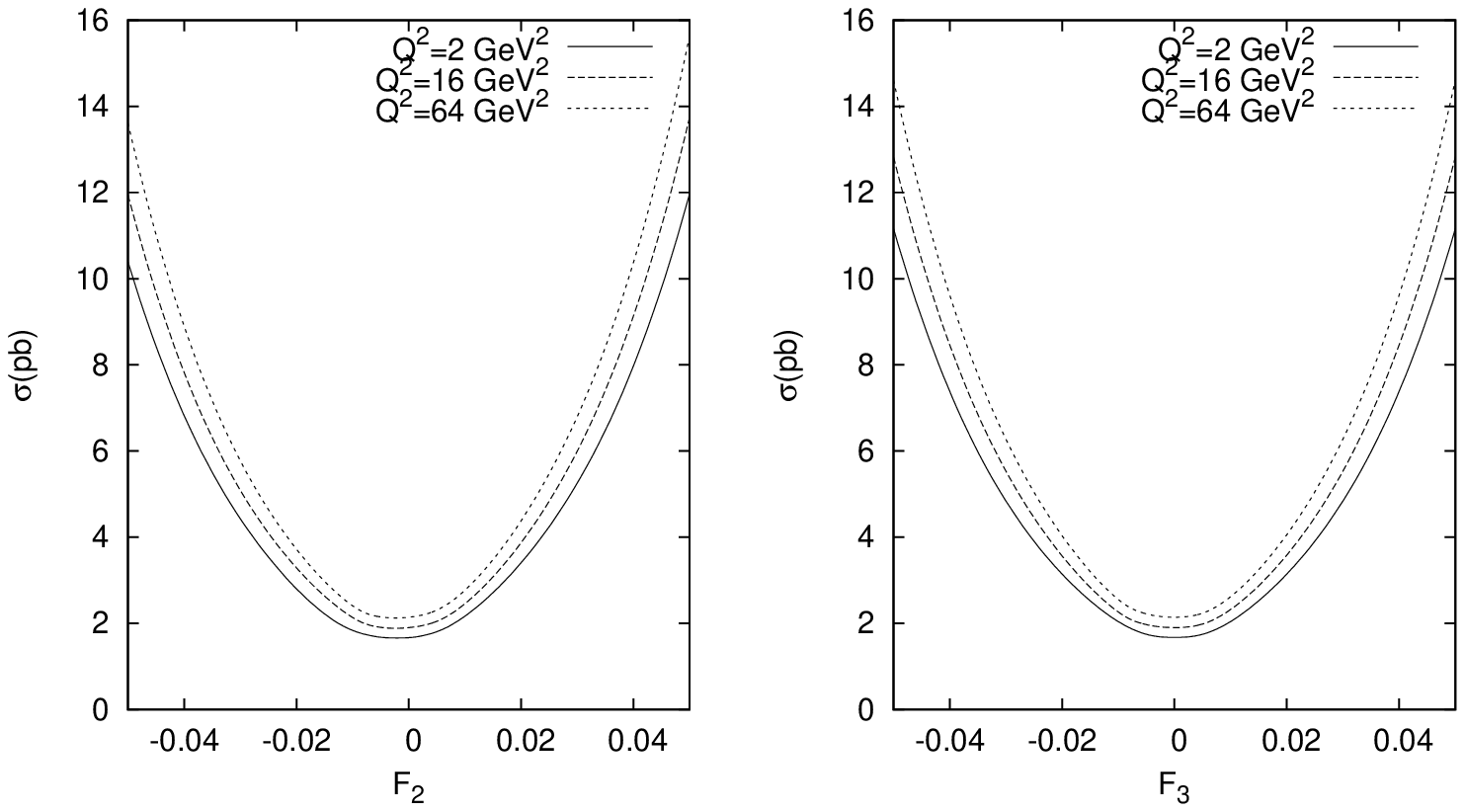}
\caption{The same as Fig. $2$ but for $\sqrt{s}=1.5$ TeV.
\label{fig3}}
\end{figure}

\begin{figure}
\includegraphics{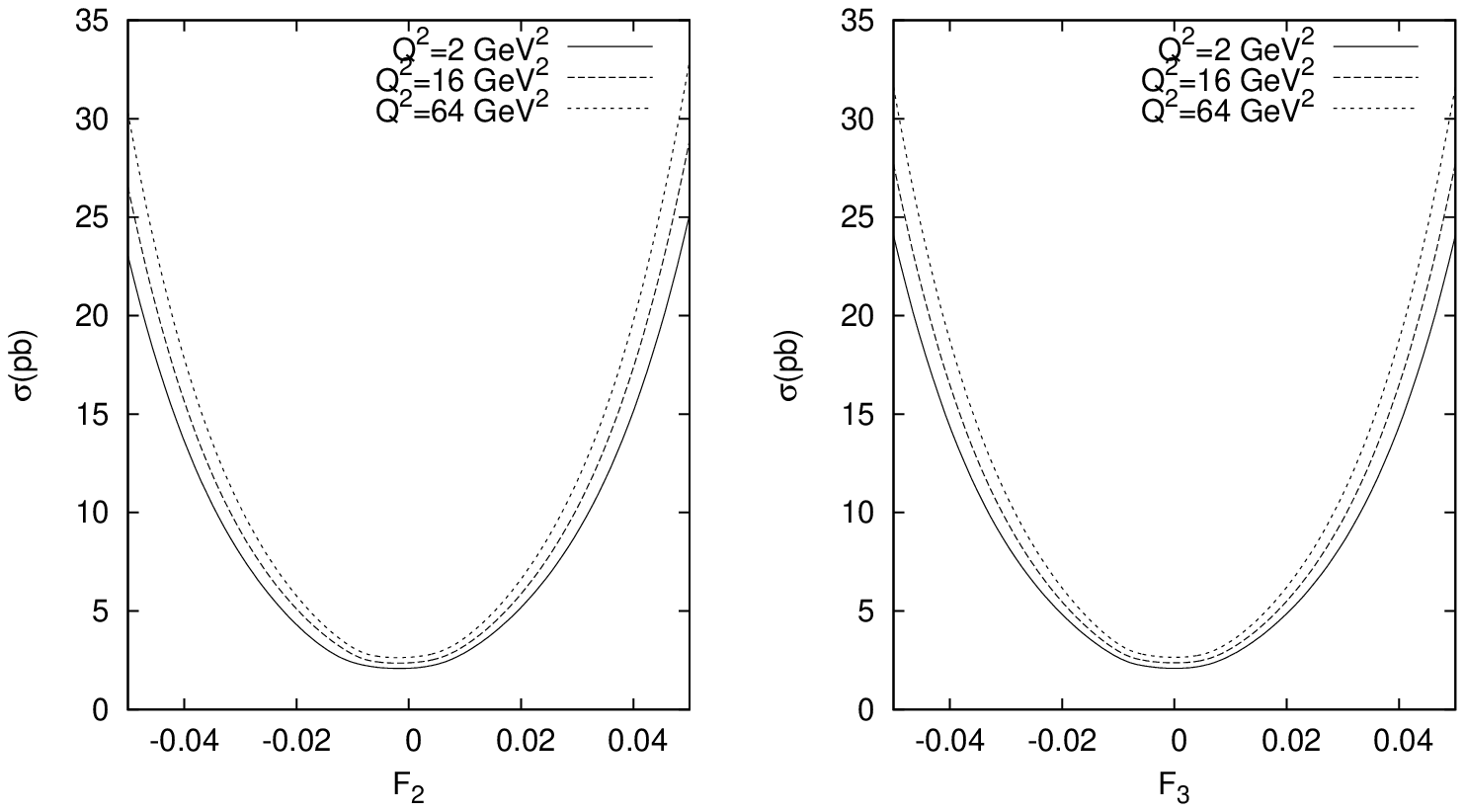}
\caption{The same as Fig. $2$ but for $\sqrt{s}=3$ TeV.
\label{fig4}}
\end{figure}

\begin{figure}
\includegraphics{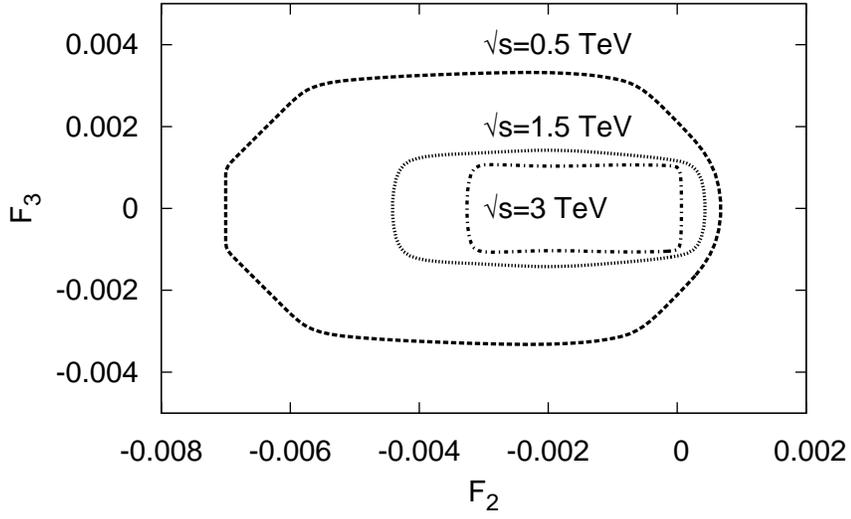}
\caption{The contour plot for the upper bounds of the anomalous couplings $F_{2}$ and $F_{3}$ with $95\%$ C.L. at the $\sqrt{s}=0.5$, $1.5$ and $3$ TeV with corresponding maximum luminosities for
the process $e^{+}e^{-} \rightarrow e^{+}\gamma^{*} \gamma^{*} e^{-}\rightarrow e^{+}\tau \bar{\tau}e^{-}$.
\label{fig5}}
\end{figure}

\begin{table}
\caption{The sensitivity limits on the anomalous couplings $a_{\tau}$ and $d_{\tau}$ for different values of photon virtuality, center-of-mass energy and luminosity.
\label{tab1}}
\begin{ruledtabular}
\begin{tabular}{ccccc}
$Q_{max}^2(GeV^2$)& $\sqrt{s}(TeV)$& Luminosity(fb$^{-1})$& $a_{\tau}$& $d_{\tau}(e,\,cm)$  \\
\hline
$2$& $0.5$& $50$& $(-0.0077, 0.0016)$  &$0.19 \times 10^{-16}$ \\
$2$& $0.5$& $230$& $(-0.0068, 0.0007)$ &$0.13 \times 10^{-16}$  \\
$2$& $3$& $200$& $(-0.0043, 0.0005)$ &$0.08 \times 10^{-16}$  \\
$2$& $3$& $590$& $(-0.0036, 0.0003)$ &$0.06\times 10^{-16}$ \\
\hline
$16$& $0.5$& $50$& $(-0.0076, 0.0015)$  &$0.19 \times 10^{-16}$ \\
$16$& $0.5$& $230$& $(-0.0067, 0.0007)$ &$0.12 \times 10^{-16}$  \\
$16$& $3$& $200$& $(-0.0042, 0.0005)$ &$0.08 \times 10^{-16}$  \\
$16$& $3$& $590$& $(-0.0036, 0.0003)$ &$0.06\times 10^{-16}$ \\
\hline
$64$& $0.5$& $50$& $(-0.0076, 0.0015)$  &$0.18 \times 10^{-16}$ \\
$64$& $0.5$& $230$& $(-0.0067, 0.0006)$ &$0.12 \times 10^{-16}$  \\
$64$& $3$& $200$& $(-0.0042, 0.0005)$ &$0.08 \times 10^{-16}$  \\
$64$& $3$& $590$& $(-0.0035, 0.0003)$ &$0.06\times 10^{-16}$ \\
\end{tabular}
\end{ruledtabular}
\end{table}

\begin{table}
\caption{95\% C.L. sensitivity bounds of the coupling $a_{\tau}$ and ${d_{\tau}}$
for various integrated CLIC luminosities and systematic uncertainties at the $\sqrt{s}=0.5$ TeV.
\label{tab2}}
\begin{ruledtabular}
\begin{tabular}{cccc}
Luminosity(fb$^{-1}$)& $\delta_{sys}$& $a_{\tau}$& $d_{\tau}(e,\,cm)$  \\
\hline
$50$& $\delta_{sys}=0$     &$(-0.0077, 0.0016)$  &$0.19 \times 10^{-16}$  \\
$50$& $\delta_{sys}=0.02$  &$(-0.0098, 0.0037)$ &$3.44 \times 10^{-16}$ \\
$50$& $\delta_{sys}=0.05$  &$(-0.0130, 0.0065)$ &$5.27 \times 10^{-16}$ \\
$50$& $\delta_{sys}=0.10$  &$(-0.0153, 0.011)$ &$7.77 \times 10^{-16}$ \\
\hline
$100$& $\delta_{sys}=0$& $(-0.0073, 0.0013)$& $0.16  \times 10^{-16}$ \\
$100$& $\delta_{sys}=0.02$& $(-0.0097, 0.0036)$& $3.33 \times 10^{-16}$ \\
$100$& $\delta_{sys}=0.05$& $(-0.0128, 0.0064)$& $5.21 \times 10^{-16}$ \\
$100$& $\delta_{sys}=0.10$& $(-0.0152, 0.011)$& $7.21 \times 10^{-16}$ \\
\hline
$230$ &$\delta_{sys}=0$     &$(-0.0068, 0.0007)$ &$0.13 \times 10^{-16}$ \\
$230$ &$\delta_{sys}=0.02$  &$(-0.0096, 0.0036)$ &$3.22 \times 10^{-16}$ \\
$230$ &$\delta_{sys}=0.05$  &$(-0.0126, 0.0062)$ &$5.10 \times 10^{-16}$ \\
$230$ &$\delta_{sys}=0.10$  &$(-0.0151, 0.010)$ &$6.66 \times 10^{-16}$ \\

\end{tabular}
\end{ruledtabular}
\end{table}

\begin{table}
\caption{95\% C.L. sensitivity bounds of the coupling $a_{\tau}$ and ${d_{\tau}}$
for integrated CLIC luminosities and various systematic uncertainties at the $\sqrt{s}=1.5$ TeV.
\label{tab3}}
\begin{ruledtabular}
\begin{tabular}{cccc}
Luminosity(fb$^{-1}$)& $\delta_{sys}$& $a_{\tau}$& $d_{\tau}(e,\,cm)$  \\
\hline
$100$& $\delta_{sys}=0$     &$(-0.0051, 0.0008)$ &$0.11\times 10^{-16}$ \\
$100$& $\delta_{sys}=0.02$  &$(-0.0076, 0.0032)$ &$2.78 \times 10^{-16}$ \\
$100$& $\delta_{sys}=0.05$  &$(-0.0102, 0.0060)$ &$4.33 \times 10^{-16}$ \\
$100$& $\delta_{sys}=0.10$  &$(-0.0132, 0.0092)$ &$6.66 \times 10^{-16}$ \\
\hline
$200$& $\delta_{sys}=0$     &$(-0.0049, 0.0006)$ &$0.10\times 10^{-16}$ \\
$200$& $\delta_{sys}=0.02$  &$(-0.0075, 0.0031)$ &$2.72 \times 10^{-16}$ \\
$200$& $\delta_{sys}=0.05$  &$(-0.0101, 0.0059)$ &$4.30 \times 10^{-16}$ \\
$200$& $\delta_{sys}=0.10$  &$(-0.0131, 0.0091)$ &$6.38 \times 10^{-16}$ \\
\hline
$320$& $\delta_{sys}=0$     &$(-0.0047, 0.0005)$ &$0.08\times 10^{-16}$ \\
$320$& $\delta_{sys}=0.02$  &$(-0.0075, 0.0030)$ &$2.66 \times 10^{-16}$ \\
$320$& $\delta_{sys}=0.05$  &$(-0.0100, 0.0058)$ &$4.27 \times 10^{-16}$ \\
$320$& $\delta_{sys}=0.10$  &$(-0.0130, 0.0090)$ &$6.11 \times 10^{-16}$ \\

\end{tabular}
\end{ruledtabular}
\end{table}

\begin{table}
\caption{95\% C.L. sensitivity bounds of the coupling $a_{\tau}$ and ${d_{\tau}}$
for integrated CLIC luminosities and various systematic uncertainties at the $\sqrt{s}=3$ TeV.
\label{tab4}}
\begin{ruledtabular}
\begin{tabular}{cccc}
Luminosity(fb$^{-1}$)& $\delta_{sys}$& $a_{\tau}$& $d_{\tau}(e,\,cm)$  \\
\hline
$200$& $\delta_{sys}=0$     &$(-0.0043, 0.0005)$ &$0.08 \times 10^{-16}$ \\
$200$& $\delta_{sys}=0.02$  &$(-0.0067, 0.0033)$ &$2.55 \times 10^{-16}$ \\
$200$& $\delta_{sys}=0.05$  &$(-0.0090, 0.0055)$ &$4.10 \times 10^{-16}$ \\
$200$& $\delta_{sys}=0.10$  &$(-0.0113, 0.0084)$ &$5.49 \times 10^{-16}$ \\
\hline
$400$& $\delta_{sys}=0$     &$(-0.0039, 0.0004)$ &$0.07 \times 10^{-16}$ \\
$400$& $\delta_{sys}=0.02$  &$(-0.0066, 0.0032)$ &$2.53 \times 10^{-16}$ \\
$400$& $\delta_{sys}=0.05$  &$(-0.0090, 0.0054)$ &$4.02 \times 10^{-16}$ \\
$400$& $\delta_{sys}=0.10$  &$(-0.0112, 0.0083)$ &$5.46 \times 10^{-16}$ \\
\hline
$590$& $\delta_{sys}=0$     &$(-0.0036, 0.0003)$ &$0.06\times 10^{-16}$ \\
$590$& $\delta_{sys}=0.02$  &$(-0.0066, 0.0032)$ &$2.50 \times 10^{-16}$ \\
$590$& $\delta_{sys}=0.05$  &$(-0.0090, 0.0054)$ &$3.99 \times 10^{-16}$ \\
$590$& $\delta_{sys}=0.10$  &$(-0.0112, 0.0082)$ &$5.42 \times 10^{-16}$ \\
\end{tabular}
\end{ruledtabular}
\end{table}

\end{document}